# Initialization Errors in Quantum Data Base Recall

Kalyani Natu
Oklahoma State University

## Abstract

This paper analyzes the relationship between initialization error and recall of a specific memory in the Grover algorithm for quantum database search. It is shown that the correct memory is obtained with high probability even when the initial state is far removed from the correct one. The analysis is done by relating the variance of error in the initial state to the recovery of the correct memory and the surprising result is obtained that the relationship between the two is essentially linear.

## Introduction

Quantum algorithms (like Shor's and Grover's) require the state of the quantum register to be precisely defined. Since any engineered system is bound to have errors, this represents what has been called the initialization problem [1] of quantum computing. In order for a quantum algorithm to run properly, the initialization error must be minimized and the influence of the residual error on the performance of the algorithm must be determined to estimate the level by which the performance will fall short of expectation.

This article considers Grover's data base search algorithm [2]-[4] for further characteristics of its performance. It has been argued that there are additional conditions that need to be kept in mind while considering its suitability [5]. Quantum data base search also has implications for how human memory works [6]-[9] and related to the walk of a quantum particle on a graph [10].

Here we determine the impact of the initialization error on the performance of the Grover algorithm for data base search [2]. We find an essentially linear relationship between the error variance associated with the initial state and the probability of finding the correct memory. The paper also sheds light on the idea of randomness in a quantum state.

## The Error Model

We assume that there is an initial vector that is able to estimate the quality of the initial sate since the actual specific state cannot be measured. One can only assume a specific error model that leads to the actual state to be different from the ideal one, which in both the Shor and the Grover algorithms should be the state where each cell of the register is in the diagonal state.

Let $a_i$ be the probability amplitudes of the elements that constitute the state. We assume that the values are real and we use the constant k to normalize the sum of the squares of the components to unit.

The initial vector matrix is designed such that the summation of the squares of all elements is unity. While for calculating the variance of the matrix, one of the necessary conditions is that the summation of all elements must also be equal to unity.

$$Variance = \sigma^2 = \frac{\sum_i (ka_i - \mu)^2}{N}$$

where

$k$ = constant which makes the summation of all elements equal to 1.

$a_i$ = data points

$\mu$ = mean of data points = $\Sigma\, ka_i / N$

$N$ = number of data points

Since the root of summation of squares of all elements in an initial vector is unity, for a N*1 initial state vector,

$\mu = k(a_1 + a_2 + a_3 + \ldots + a_N) / N,$

$\sqrt{(a_1^2 + a_2^2 + a_3^2 + \ldots + a_N^2)} = 1$

Hence the variance will be, $\sigma^2 = 1/N\, \Sigma\, (ka_i - \Sigma\, ka_i/N)^2$     (1)

$$= k^2/N\, (a_i - \mu)^2$$

$$= k^2\, \{1/N - 1/N^2\, (a_1 + a_2 + a_3 + \ldots + a_N)^2\}$$

To maximize the variance, $(a_1 + a_2 + a_3 + \ldots + a_N)^2/N^2 = 0$, hence the maximum variance

$$\sigma_{max}^2 = k^2/N \quad\quad\quad\quad (2)$$

The maximum value of k will be 1 and hence $\sigma_{max}^2 = 1/N$.

The above equation holds true only if the elements of the initial vector are real numbers.

Our simulations corroborate that the variance increases or approaches the maximum value, the probability decreases and reaches a minimum of about 50% near $\sigma_{max}^2$.

We now use Markov's inequality, according t which, if X is a nonnegative random variable and a > 0, then

$$P(X \geq a) \leq E(X)/a \qquad (3)$$

Using this equation for the variance of initial vector matrix, since, $\Sigma |a_i|^2 = 1$, therefore, $E[|a_i|^2] = 1\backslash N$.

We can now write,

$$P[|a_i|^2 \geq 1/N + €] \leq E[x] / (1/N) + €$$

$$\leq (1/N) / (1/N + €)$$

$$\leq 1 + 1/N€ \qquad (4)$$

## Database search

Consider N unsorted items in a database, out of which just one item is to be retrieved. By reviewing the item, it is easily possible to guess whether the item satisfies the required condition or it is the exact item which is to be looked for. The main drawback of the search is that the database is unsorted and hence there is no easier way in classical algorithm to find the specific item than by reviewing each item one at a time. Due to its unorderly state, the user has to keep track of the items reviewed and proceed to next item in the database so that the same item is not reviewed again. Therefore, the user will have to examine around N/2 items before reaching the looked-for item. The classical algorithm will take O(N) steps to reach to the desired item.

Grover's algorithm requires $O(\sqrt{N})$ steps. A system has $N=2^n$ ($S_1, S_2,\ldots,S_n$) states out of which only one state satisfies the condition $C(S_v)=1$, while all others states have $C(S)=0$. The problem is to search for the state $S_v$. The algorithm works as follows:

Step 1: Set up the problem.

Initialize the system to the distribution $(1/\sqrt{N}, 1/\sqrt{N},\ldots,1/\sqrt{N}) = (a_1,a_2,\ldots,a_N)$ where the amplitude is same for each of the N states and the root of summation of the squares of all elements is unity. This is the Initial vector, which essentially is a device having the ability to recognize the solutions to search problems. Initial state of initial vector is |0>.

Step 2: Repeat the unitary operations $O(\sqrt{N})$ times, i.e. to reach the solution $O(\sqrt{N})$ iterations are required.

a) If C(S)=1, rotate phase by $\pi$ radians, if C(S)=0, system is unchanged.
b) Apply diffusion transform defined by: $D_{ij} = 2/N$ if $i \neq j$ & $D_{ii} = -1+2/N$.
c) Flip sign of item to be searched for and measure the state.

Step 3: Measure the state, in case $C(S_v)=1$, there is a unique state $S_v$. If not, repeat the above steps for $O(\sqrt{N})$ times.

The probability of finding the correct item is for various sizes of the problem is as follows:

| Database size | Probability of finding correct item |
|---|---|
| 4 | 100 |
| 8 | 97.22 |
| 16 | 98.02 |
| 32 | 99.95 |

## Initial Vector with Errors

Grover algorithm has huge potential benefits when searching very large, disorderly data sets. But in practical systems, the initial vector will not be as accurate as defined in the algorithm. There will always be dissimilarities between all elements, although the condition of root of sum of squares of all elements being one will be preserved.

Suppose that an initial vector is generated randomly and then a specific item is to be searched for in a data base then how would that affect the probability of finding the item. This means that instead of having the initial state as |0>, if we start with an arbitrary initial state then how is the probability affected.

For an N=8 and initial vector with arbitrary initial state, **N = $2^3$ = 8**, there are **3 qubits** and the number of iterations = $\sqrt{N} = \sqrt{8} = 2.82 \approx 3$. The same steps as the Grover algorithm are followed. A diffusion matrix is generated by the above mentioned formula. It looks like this.

| -0.7500 | 0.2500 | 0.2500 | 0.2500 | 0.2500 | 0.2500 | 0.2500 | 0.2500 |
|---|---|---|---|---|---|---|---|
| 0.2500 | -0.7500 | 0.2500 | 0.2500 | 0.2500 | 0.2500 | 0.2500 | 0.2500 |
| 0.2500 | 0.2500 | -0.7500 | 0.2500 | 0.2500 | 0.2500 | 0.2500 | 0.2500 |
| 0.2500 | 0.2500 | 0.2500 | -0.7500 | 0.2500 | 0.2500 | 0.2500 | 0.2500 |
| 0.2500 | 0.2500 | 0.2500 | 0.2500 | -0.7500 | 0.2500 | 0.2500 | 0.2500 |
| 0.2500 | 0.2500 | 0.2500 | 0.2500 | 0.2500 | -0.7500 | 0.2500 | 0.2500 |
| 0.2500 | 0.2500 | 0.2500 | 0.2500 | 0.2500 | 0.2500 | -0.7500 | 0.2500 |
| 0.2500 | 0.2500 | 0.2500 | 0.2500 | 0.2500 | 0.2500 | 0.2500 | -0.7500 |

Fig 1: 8*8 diffusion matrix

A random initial vector matrix is generated and then normalized so that the condition of sum of squares of all elements is unity is satisfied. The initial vector matrix is different every time the code is run. The code is designed such that we are trying to search for the second element and after every time the Grover algorithm is terminated, the probability of getting the correct solution is measured. To measure the errors in the initial vector, the property called variance is measured over the initial vector. The final plot is the probability vs the measure of variance.

The plots vary every time due to the randomness of the initial vector matrix and hence the measure of variance. The results obtained are as follows:

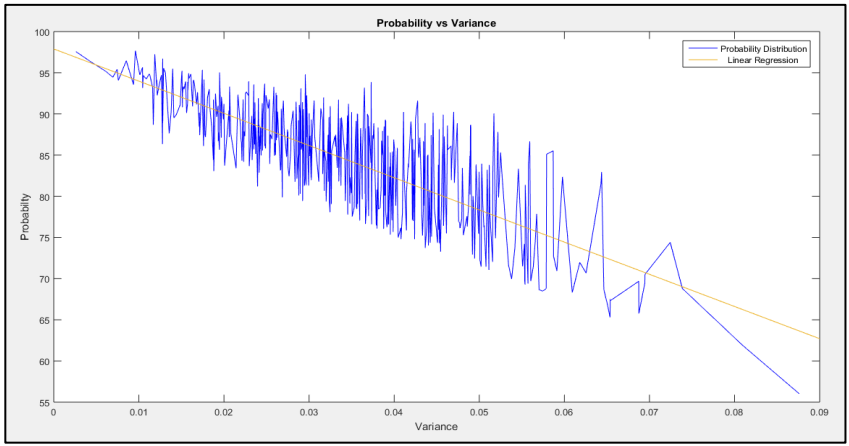

Fig 2: Probability Distribution for 8*1 initial vector matrix (1)

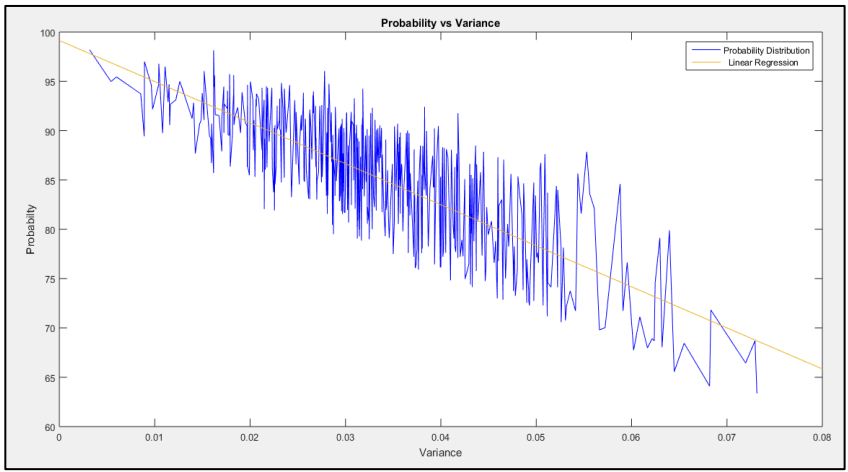

Fig 3: Probability Distribution for 8*1 initial vector matrix

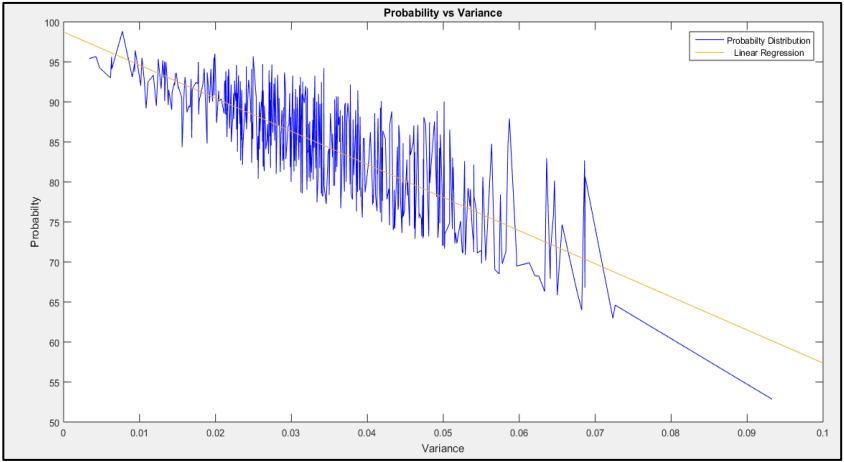

Fig 4: Probability Distribution for 8*1 initial vector matrix

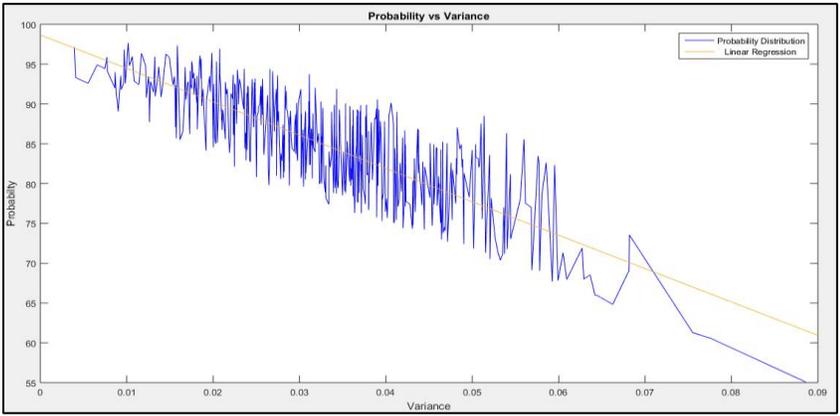

Fig 5: Probability Distribution for 8*1 initial vector matrix

By calculations, if the initial vector is such that all elements are similar, then the probability of finding the required item is 97.22%. The above plots demonstrate how the probability of finding the item decreases as the variance i.e. the error component in the initial vector increases.

For N=16 and an initial vector with arbitrary initial state, $N = 2^4 = 16$, there are 4 qubits and the number of iterations = $\sqrt{N} = \sqrt{16} = 4$. The same steps as the Grover algorithm are followed.

The plots of probability vs measure of variance is as follows:

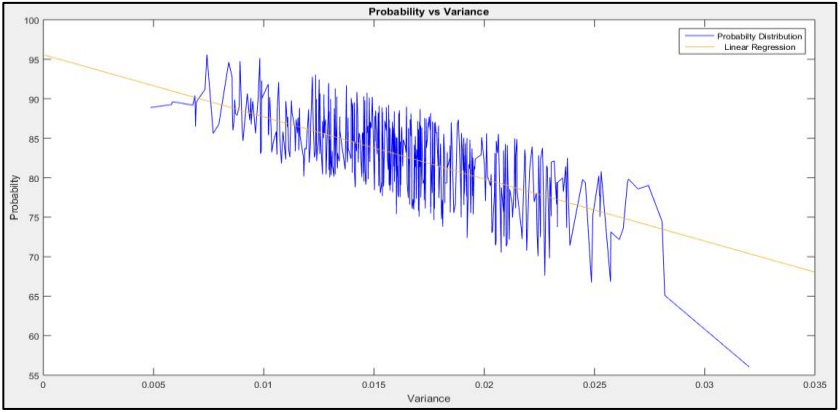

Fig 6: Probability Distribution for 16*1 initial vector matrix (1)

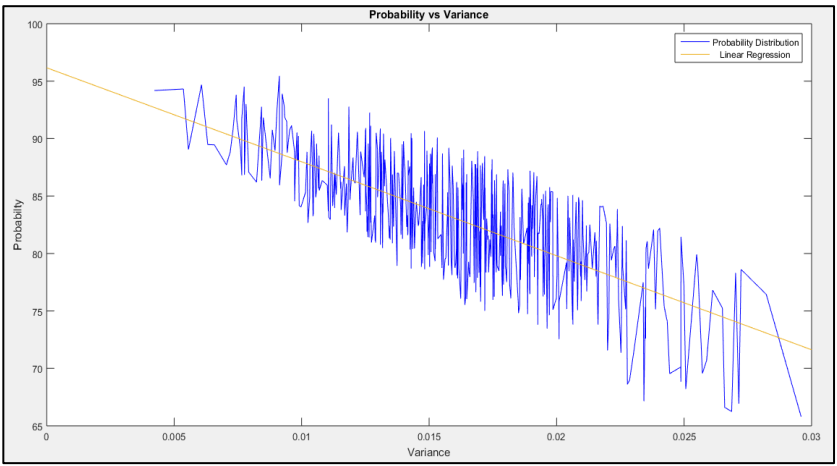

Fig 7: Probability Distribution for 16*1 initial vector matrix

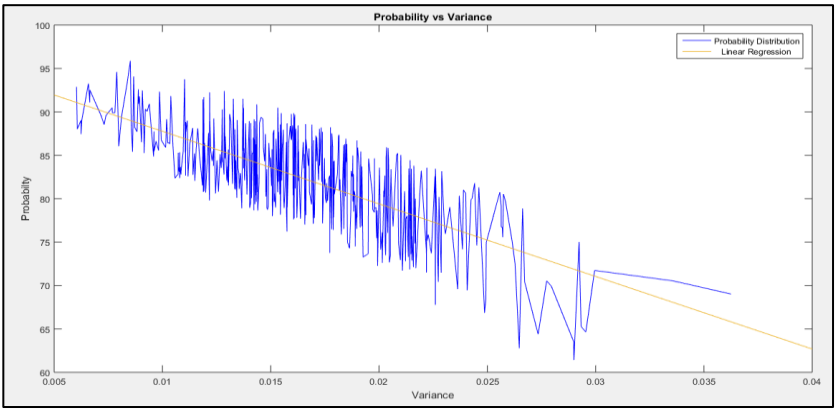

Fig 8: Probability Distribution for 16*1 initial vector matrix

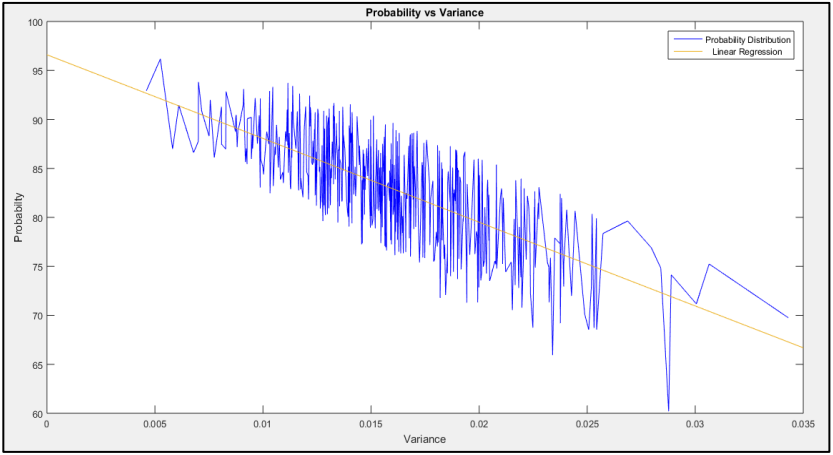

Fig 9: Probability Distribution for 16*1 initial vector matrix

By calculations, if the initial vector is such that all elements are similar, then the probability of finding the required item is 98.02%. The above plots demonstrate how the probability of finding the item decreases as the variance i.e. the error component in the initial vector increases.

For a 4*4 diffusion matrix, initial vector is one with arbitrary initial state. Since $N = 2^{2} = 4$, there are 2 qubits and the number of iterations = $\sqrt{N} = \sqrt{4} = 2$. The same steps as the Grover algorithm are followed. A diffusion matrix is generated by the above mentioned formula.

The plot of probability vs measure of variance is as follows and as we see the relationship between probability and error is essentially linear.

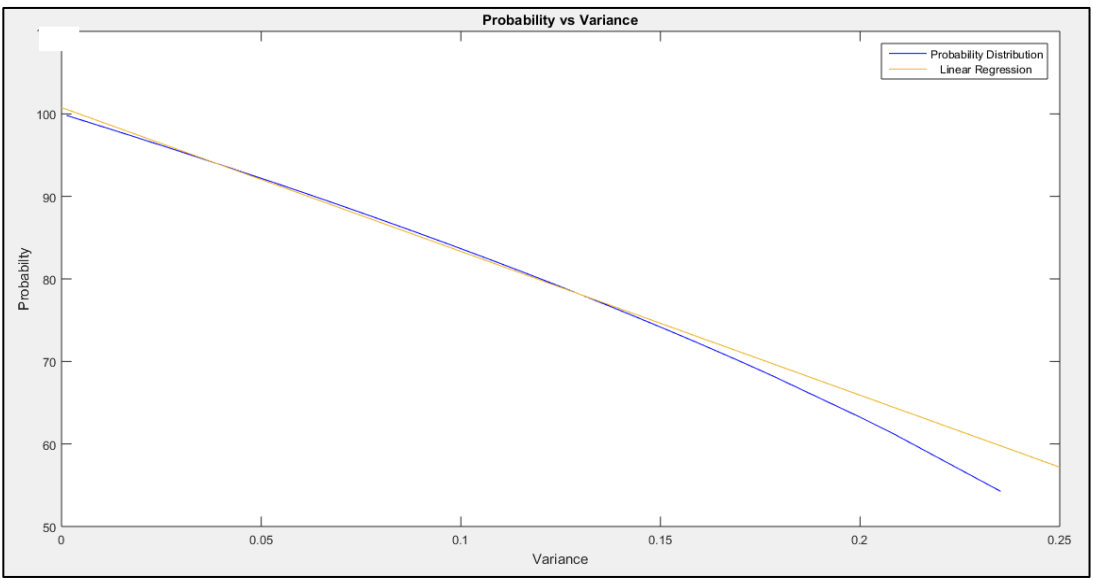

Fig 10: Probability Distribution for 4*1 initial vector matrix (3)

The above plot demonstrates how the probability of finding the item decreases as the variance i.e. the error component in the initial vector increases.

## Conclusion

The research in this paper provides shows how the probability of finding the item in an unsorted database decreases as the errors in the initial vector matrix increases. Since the probability doesn't decrease below 50% this algorithm could be relevant to find items in unsorted and very large databases even when the initial state is not error free. This work can be further extended to find out how the probability gets affected if the elements of initial vector are complex or if there are other errors in the diffusion matrix.